\documentclass[prb,amsmath,amssymb,twocolumn]{revtex4}

\usepackage{graphicx}
\usepackage{subfigure}
\usepackage{adjustbox}
\usepackage{bm}
\usepackage{color}
\usepackage{braket}
\usepackage{standalone}
\usepackage{multirow}
\usepackage{tikz}
\usepackage{mathrsfs}
\usepackage[colorlinks,bookmarks=true,citecolor=blue,linkcolor=red,urlcolor=blue]{hyperref}

\bibpunct{[}{]}{,}{n}{}{}

\begin{document}

\title{Extended Bloch theorem for topological lattice models with open boundaries }

\author{Flore K. Kunst,$^{1}$ Guido van Miert$^{2}$ and Emil J. Bergholtz$^{1}$}

\affiliation{$^1$ Department of Physics, Stockholm University, AlbaNova University Center, 106 91 Stockholm, Sweden
\\$^2$ Institute for Theoretical Physics, Centre for Extreme Matter and Emergent Phenomena, Utrecht University, Princetonplein 5, 3584 CC Utrecht, The Netherlands}
\date{\today}

\begin{abstract}
While the Bloch spectrum of translationally invariant noninteracting lattice models is trivially obtained by a Fourier transformation, diagonalizing the same problem in the presence of open boundary conditions is typically only possible numerically or in idealized limits. Here we present exact analytic solutions for the boundary states in a number of lattice models of current interest, including nodal-line semimetals on a hyperhoneycomb lattice, spin-orbit coupled graphene, and three-dimensional topological insulators on a diamond lattice, for which no previous exact finite-size solutions are available in the literature. Furthermore, we identify spectral mirror symmetry as the key criterium for analytically obtaining the entire (bulk and boundary) spectrum as well as the concomitant eigenstates, and exemplify this for Chern and $\mathcal Z_2$ insulators with open boundaries of co-dimension one. In the case of the two-dimensional Lieb lattice, we extend this further and show how to analytically obtain the entire spectrum in the presence of open boundaries in both directions, where it has a clear interpretation in terms of bulk, edge, and corner states.
\end{abstract}

\maketitle

\section{Introduction}

One of the main subjects of investigation in the condensed matter community are topological phases of matter, which already in the single-particle limit offer a plethora of fascinating properties such as the appearance of dissipationless, robust boundary states \cite{hasankane, qizhang, Weylreview}. The boundary modes are not only a signature feature of first-order topological phases---e.g., Chern insulators, $\mathcal{Z}_2$ insulators, and strong, three-dimensional topological insulators feature chiral edge modes, helical edge modes, and two-dimensional Dirac states, respectively, on their boundaries \cite{klitzingdordapepper, haldane, tknn, hofstadter, changzhangfengshenzhang, jotzumesserdesbuquoislebratuehlingergreifesslinger, kaneandmele, kaneandmeletwo, bernevigzhang, bernevighugheszhang, koenigwiedmannbruenerothbuhmannmolenkamp, kanemelestrongti, rhimbardarsonslager}---higher-order topological phases are also marked by the appearance of boundary modes, which now localize to terminations with a codimension larger than one \cite{sittefoschaltmanfritz, benalcazarbernevighughes, langhehnpentrifuoppenbrouwer, linhughes, songfangfang, benalcazarbernevighughesagain, schindlercookvergnio, geiertrifunovichoskambrouwer, trifunovicbrouwer, vanmiertortix, kooivanmiertortix, kunstvmiertbergholtz, kunstvmiertbergholtz2, vanmiertortixmoraissmith, vanmiertortix2, slagerrademakerzaanenbalents}.

\begin{figure}[b]
  \centering
  {\includegraphics[scale=0.28]{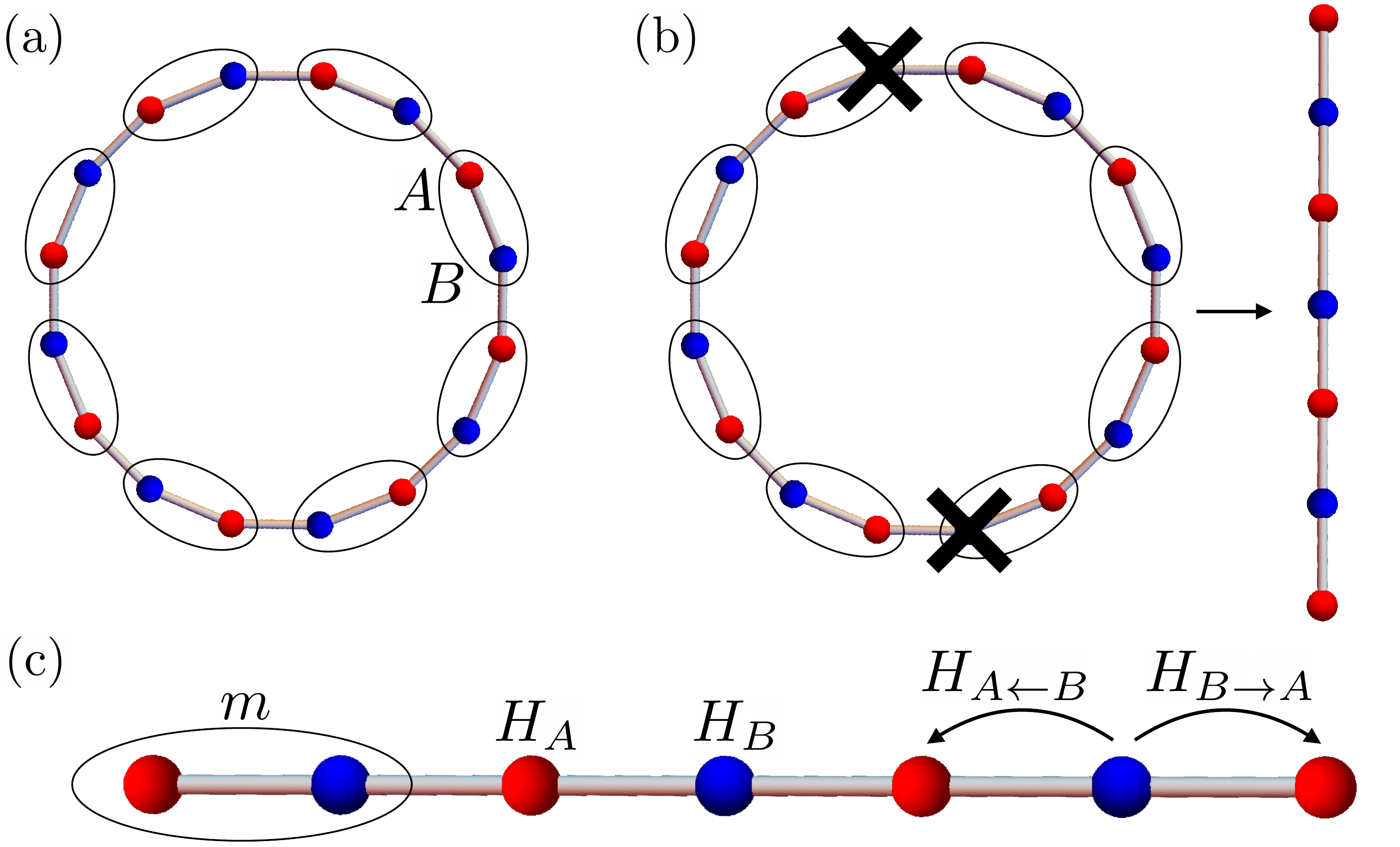}}
  \caption{(a) Schematic depiction of periodic lattice consisting of $A$ and $B$ motifs in red and blue, respectively, with each unit cell encircled by an ellipse. (b) By removing the $B$ site from the $0$th (or $2M$th) and $M$th unit cell, one obtains an open chain with $A$ motifs at its ends. (c) The Hamiltonian for the open chain with each unit cell labeled by $m$.}
    \label{figurechainmodels}
\end{figure}

To access these boundary modes, one needs to consider a system with open boundary conditions, which greatly enlarges the unit cell and consequently renders the problem of finding analytical solutions more challenging and generally intractable in the limit of large system sizes. Despite this increased difficulty, we showed in previous works that exact solutions for the boundary-mode wave functions of any codimension $d$ can be straightforwardly obtained on a large family of $D$-dimensional lattices by making use of the naturally present destructive interference and without the need for fine-tuned tight-binding parameters \cite{kunsttrescherbergholtz, kunstvmiertbergholtz, kunstvmiertbergholtz2}. The lattices we investigate consist of repeated structures that are subdivided in $A$ and $B$-type motifs, and have $A$ motifs at their boundaries, e.g., Fig.~\ref{figurechainmodels}(c). On these lattices, we find $n$ exact solutions, where $n$ is the number of degrees of freedom of the $A$ motifs, which exponentially localize to the boundaries with a completely disappearing amplitude on the $B$-type motifs and whose eigenenergies correspond to the eigenenergies of the local Hamiltonian on $A$. Other methods also exist \cite{leejoannopoulos, hatsugai, dwivedichua, alasecobaneraortizviola, cobaneraalaseortizviola, duncanoehbergvaliente, lopezsanchotworubio, umerski}, which are restricted to finding solutions for models with open boundary conditions in one direction only, while a generalization to higher-order phases is also made in an approximate fashion in Ref.~\onlinecite{pengbaovonoppen}. Despite this recent progress, there are several important questions regarding the exact treatment of the eigensystem of lattice models with open boundary conditions that remain unanswered, and we here systematically address them building on our framework introduced in Refs.~\onlinecite{kunsttrescherbergholtz, kunstvmiertbergholtz, kunstvmiertbergholtz2}.

Firstly, we investigate the premises for when it is possible to obtain, in addition to recovering boundary states, all remaining bulk states, and we find that this is the case when there is a mirror symmetry in the spectrum of the periodic Bloch Hamiltonian, $E(\vec{k}_{||}, k_\perp) = E(\vec{k}_{||}, - k_\perp)$, where the mirror relates the momenta $k_\perp$ in the direction of the open boundary conditions. In the presence of this symmetry, the problem of finding the complete eigensystem reduces to the problem of solving for the standing waves in an open string. Using this method, we are also able to show that the boundary wave functions that we have previously recovered are the only boundary modes in the system. We also provide a case study of the Lieb lattice where we obtain the full spectrum with open boundary conditions in both directions, where the solutions have a particularly illuminating interpretation in terms of bulk, edge, and corner states, respectively.

This leads to the second question, which addresses the issue of completeness of our boundary eigensystem, namely, in the absence of a spectral mirror symmetry, are the boundary-mode solutions we have found unique, or are there other boundary modes in the system that we have not found? This conundrum is resolved upon using a result derived using transfer matrices regarding the structure of edge states, which explicitly allows us to show that the boundary-mode solutions we find are the only solutions that exist.

Finally we note that certain mechanisms such as spin-orbit coupling (SOC) have have not been accounted for in previous studies. Here we investigate when it is possible to still solve for the boundary eigensystem, or the entire eigensystem in case of spectral mirror symmetry, in the presence of SOC. We find that as long as the spin on the $A$ motifs is well defined, and the coupling between different $A$ and $B$ motifs is invariant under spin rotation, we can still use our formalism to find analytical solutions. This is a particularly interesting problem to address in the case of three-dimensional strong topological insulators, where the spin sectors need to be explicitly coupled via SOC to obtain Dirac surface states.

We illustrate our findings by introducing explicit examples. To showcase our previously developed method, we introduce a ``vanilla" example, which does not have a spectral mirror symmetry nor includes SOC, by considering a simple nearest-neighbor hopping model on the hyperhoneycomb lattice \cite{mullenuchoaglatzhofer, mullenuchoawangglatzhofer, ezawahyper}, which is the natural structure of $\beta$-Li$_2$IrO$_3$ \cite{modicsmidtkimchibreznayetal}. This model realizes a nodal-line semimetal, which are systems that feature a degeneracy in the spectrum that forms a closed curve that is explicitly protected by the presence of spatial symmetries or time-reversal symmetry \cite{burkovhookbalents}. We find that our solution coincides with the wave function of the drumhead surface state, which corresponds to an exactly flat band in the spectrum. We thus solve for the drumhead state in a nodal-line semimetal in an exact fashion.

We continue by considering a variation to the Haldane model on the honeycomb lattice \cite{haldane}, which realizes a Chern-insulating phase and has the desired spectral mirror symmetry. We show that we are indeed not only able to find solutions to the chiral edge states but also solve for the complete eigensystem. Next, we introduce spin degree of freedom in this model by considering two time-reversed copies of the pseudo-Haldane model by which we realize a $\mathcal{Z}_2$ insulator in the form of a pseudo-Kane-Mele model \cite{kaneandmele,kaneandmeletwo}. This expansion of our Hilbert space does not spoil the spectral mirror symmetry such that we can still solve for the full eigensystem with our boundary-mode solutions now corresponding to the helical edge states. For this model, the SOC does not connect the two spin sectors, and it is thus straight forward to solve this model.

However, we note that while the strategy of gluing two Chern insulators with opposite Chern number together indeed leads to the retrieval of a $\mathcal{Z}_2$ insulator, this approach does not suffice to obtain a strong topological insulator in three dimensions and we need to explicitly include a SOC, which connects the different spin sectors, to accomplish this phase. We implement a variation to the model introduced by Fu \emph{et al.} \cite{kanemelestrongti}, which realizes a strong topological insulator on the diamond lattice, and we show that we can find the boundary states both in the topologically trivial phase, the topologically weak phase, and the topologically strong phase. Moreover, we connect this phase to a nodal-line semimetal phase by which a direct connection between Dirac surface states and drumhead states is established.

In a final case study, we dissect the band spectrum of the two-dimensional Lieb lattice with open boundary conditions in two directions. By making use of the Bloch Hamiltonian as well as the two Hamiltonians relating to the two different types of edges that appear, we are able to solve for the entire spectrum analytically.

This paper is ordered as follows. We introduce our general method in Sec.~\ref{sec:exactsol}, where we summarize our formalism in Refs.~\cite{kunsttrescherbergholtz, kunstvmiertbergholtz, kunstvmiertbergholtz2} in Sec.~\ref{sec:exactsolbounds} for the sake of completion. In Sec.~\ref{sect:fulldiag}, we show that the full eigensystem can be found in the presence of a spectral mirror symmetry, in Sec.~\ref{sec:uniqueness} we address the completeness of our boundary mode solutions, and we consider the case of SOC in Sec.~\ref{sec:soc}. The examples are introduced in Sec.~\ref{sec:examples}, and the discussion can be found in Sec.~\ref{sec:discussion}.

\section{Exact solutions} \label{sec:exactsol}

In this section, we study lattices with open boundary conditions in one direction of the form shown in Fig.~\ref{figurechainmodels}(c), and show that in addition to finding exact solutions for the boundary modes \cite{kunsttrescherbergholtz, kunstvmiertbergholtz, kunstvmiertbergholtz2}, we can also fully diagonalize the entire system if the spectral symmetry $E(\vec{k}_{||}, k_\perp) = E(\vec{k}_{||}, - k_\perp)$ is present in the spectrum of the periodic Bloch Hamiltonian. Moreover, we show that this method still works in the presence of SOC by satisfying weak constraints.

We start by considering a $D$-dimensional lattice with $M$ unit cells, which consists of two distinguishable $(D-1)$-dimensional motifs, $A$ [red in Fig.~\ref{figurechainmodels}(c)] and $B$ (blue). The $A$ motifs have $n$ degrees of freedom while the $B$ motifs have a single degree of freedom, and electrons are forbidden to hop between $A$ motifs directly. For simplicity, we set $D=1$ in the following discussion but note that the formalism can easily be generalized to higher dimensions by dimensional extension. We assume that hopping between $A$ and $B$ motifs takes place in a nearest-neighbor fashion, such that the real-space Hamiltonian reads
\begin{align}
\hat{H}&=\sum_{m,i,i'} c^\dagger_{A_{i},m}H^{i,i'}_A c_{A_{i'},m} + \sum_m H_B c^\dagger_{B,m} c_{B,m} \nonumber\\
&+\sum_{m,i}c^\dagger_{A_{i},m}H_{A\leftarrow B}^i c_{B,m}+H.c. \nonumber \\
&+\sum_{m,i}c^\dagger_{A_{i},m+1}H_{B\rightarrow A}^i c_{B,m}+H.c., \label{eqrealspaceham}
\end{align}
where $c^\dagger_{A_i, m}$ ($c^\dagger_{B,m}$) creates an electron in the $m$th unit cell within motif $A$ ($B$), and $i$, $i'$ corresponds to the internal degrees of freedom on the $A$ motif, which may be a sublattice or orbital degree of freedom. The Hamiltonian is depicted in Fig.~\ref{figurechainmodels}(c).

\subsection{Boundary modes} \label{sec:exactsolbounds}

In Refs.~\onlinecite{kunsttrescherbergholtz, kunstvmiertbergholtz, kunstvmiertbergholtz2} we showed that if these lattices terminate with $A$ motifs on either side [cf. Fig.~\ref{figurechainmodels}(c)], the following $n$ solutions exist:
\begin{equation}
\ket{\psi_i} = \mathcal{N}_i \sum_{m=1}^M r_i^m c^\dagger_{A_{i},m}\ket{0},\label{eqexactsolution}
\end{equation}
whose eigenvalues correspond to the eigenvalues $E_i$ of the Hamiltonian $H_A$ on motif $A$, and where $i$ labels the degrees of freedom on each $A$ motif, $\mathcal{N}_i$ is the normalization constants, and $c^\dagger_{A_{i},m}$ creates an electron on motif $A$ in unit cell $m$ with energy $E_i$. This wave function has zero weight on each of the $B$ motifs due to exact destructive interference, and we find
\begin{align}
r_i=-\frac{\left(H_{A\leftarrow B}^\dagger\right)^{1,i}}{\left(H_{B\rightarrow A}^\dagger\right)^{1,i}}. \label{eqexactsolr}
\end{align}
Note that these simplified forms of Eqs.~(\ref{eqexactsolution}) and (\ref{eqexactsolr}) assume that $H_A$ is written in its diagonal basis. When this is not the case, the exact wave functions take the following more general form:
\begin{equation}
\ket{\psi_i} = \mathcal{N}_i \sum_{m=1}^M r_i^m \left(\sum_{\bar{n}=1}^n c^\dagger_{A_{\bar{n}},m} \phi_{i,\bar{n}}\right) \ket{0} \label{eqexactsolrnondiag}
\end{equation}
with
\begin{equation}
r_i = -\frac{H_{A\leftarrow B}^\dagger \phi_i}{H_{B\rightarrow A}^\dagger \phi_i}, \label{eqexactsolrextra}
\end{equation}
where $\phi_{i,\bar{n}}$ is the $\bar{n}$th component of the $i$th eigenfunctions $\phi_i$ of $H_A$. Note that $r_i$ is a scalar since $H_{A\leftarrow B}^\dagger$ and $H_{B\rightarrow A}^\dagger$ are $1\times n$ matrices and $\phi_i$ is an $n$-component vector. The state $\ket{\psi_i}$ corresponds to a bulk mode when $|r_i|=1$, while localizing to the $A$ motif $m=1$ ($M$) when $|r_i|<1$ ($>1$). In the latter case, these solutions thus correspond to boundary states.

If the lattice instead terminates with an $A$ motif on one end and a $B$ motif on the other, we find that our solution in Eq.~(\ref{eqexactsolution}) is still relevant. Indeed, in this case, boundary modes appear on both ends of the lattice and when the lattice is sufficiently long, the solution in Eq.~(\ref{eqexactsolution}) may be mapped onto either end of the lattice as long as $|r_i|<1$. In this case, the solution on the boundary terminating with an $A$ ($B$) motif decays exponentially into the bulk and has an approximate zero weight on the $B$ ($A$) motifs. When $|r_i|>1$, on the other hand, the boundary states interfere with each other, such that the wave function in Eq.~(\ref{eqexactsolution}) is no longer a good approximation. Nevertheless, $|r_i|=1$ predicts when the boundary states enter the bulk and gap out up to finite size corrections and thus also provides key information for open systems without broken unit cells.

We emphasize that this method can also be used to find the edge and surface states of two- and three-dimensional crystals, respectively, as they can be seen as a collection of one-dimensional Hamiltonians parametrized by $\vec{k}_{||}$. In these cases, the function $r_i(\vec{k}_{||})$ is momentum dependent, which increases the likelihood of the states to localize to the edges or surfaces. Indeed, by setting the dimension $D=2$ and $3$ and implementing the relevant lattice Hamiltonian, we showed in Ref.~\onlinecite{kunsttrescherbergholtz} that these solutions correspond to the chiral, edge modes of a Chern insulator and the Fermi arcs in Weyl semimetals, respectively. To showcase this formalism explicitly, we show an explicit example in Sec.~\ref{sect:nodallinesm} of a three-dimensional nodal-line semimetal, where the drumhead state is captured by our exact solutions. 

We point out that in Refs.~\onlinecite{kunstvmiertbergholtz, kunstvmiertbergholtz2}, we generalized our formalism to higher-order boundaries by considering lattices that consist of a superposition of the chains like the one in Fig.~\ref{figurechainmodels}(c) into an arrangement on which solutions of the form of $\ket{\psi_i}$ exist with additional prefactors $r'_i, r''_i, \ldots$, which were found to capture corner modes as well as chiral hinge states. Indeed, in Sec.~\ref{sec:lieb_lattice} we show an explicit example of this generalization.

\subsection{Complete diagonalization of mirror-symmetric lattice models} \label{sect:fulldiag}

So far, we have only looked at retrieving boundary-mode solutions. In this section, we show we can find \emph{all} energy eigenvalues and wave functions in the presence of the spectral symmetry $E(k_\perp) = E(-k_\perp)$ of the Bloch Hamiltonian. We start by considering a periodic chain with $2M$ unit cells, where each unit cell consists of $A$ and $B$ motifs as shown in Fig.~\ref{figurechainmodels}(a). We find the Bloch Hamiltonian for this model by Fourier transforming the creation and annihilation operators in the real-space Hamiltonian in Eq.~(\ref{eqrealspaceham})
\begin{align}
c^\dagger_{A_i, k_\perp}&= \frac{1}{\sqrt{2M}}\sum_m e^{ik_\perp m}c^\dagger_{A_i, m},\nonumber\\
c^\dagger_{B, k_\perp}&= \frac{1}{\sqrt{2M}} \sum_m e^{ik_\perp m} c^\dagger_{B, m},
\end{align}
such that $\hat{H}(k_\perp)=(c^\dagger_{A_i, k_\perp},c^\dagger_{B, k_\perp})H(k_\perp)(c_{A_i, k_\perp},c_{B, k_\perp})^T $ with
\begin{align}
H(k_\perp)&=\begin{pmatrix}
H_A&H_{A\leftarrow B}+H_{B\rightarrow A}e^{-ik_\perp}\\
H^\dagger_{A\leftarrow B} +H^\dagger_{B\rightarrow A}e^{ik_\perp}&H_B
\end{pmatrix}. \label{eqgeneralblochham}
\end{align}
We denote the eigenvectors of $H(k_\perp)$ with $\Psi_{l}(k_\perp)$, where $l=1,\ldots,n+1$ is the band index. The corresponding real-space wave function in the $m$th unit cell is then given by
\begin{align}
\Psi_{l}(k_\perp,m):=e^{ik_\perp m}\Psi_{l}(k_\perp).
\end{align}

We proceed by cutting the chain in half by removing the $B$ sites in the $0$th (or $2M$th) and $M$th unit cell, such that two open chains remain with each $M$ A motifs and $(M-1)$ $B$ sites as shown in Fig.~\ref{figurechainmodels}(b). Analogously to the standing waves in an open string, almost all wave functions for the geometry of such a chain can be obtained. If we now imagine that the $B$ sites at the end of the chain are reattached, and we suppose that the wave functions $\Psi_{l}(k_{\perp, j},m)$ and $\Psi_{l}(-k_{\perp, j},m)$ with $k_{\perp, j}=\pi j/M, \, j=1,\ldots,M-1$ have the same energy, i.e., $E_l(k_\perp) = E_l(-k_\perp)$, we can form a new eigenstate by considering the following linear combination
\begin{align*}
\Psi_{{\rm Bulk},l,i} (k_{\perp, j},m) = \mathcal{A} \,  \Psi_{l,i}(k_{\perp, j},m) + \mathcal{B} \, \Psi_{l,i}(-k_{\perp, j},m) \\
= \mathcal{A} \, e^{ik_{\perp, j} m}  \Psi_{l,i}(k_{\perp, j}) + \mathcal{B} \, e^{-ik_{\perp, j} m} \Psi_{l,i}(-k_{\perp, j}),
\end{align*}
where $i = 1, \ldots, n+1$ refers to the component of the $l$th Bloch wave $\Psi_{l}(k_\perp)$. Imposing that this linear combination, the so-called ``generalized bulk state" $\Psi_{{\rm Bulk},l,i} (k_{\perp, j},m)$, vanishes on the ``reattached" zeroth and $M$th $B$ motifs, we find, both for $m = 0$ and $m=M$, that
\begin{align*}
&\Psi_{{\rm Bulk},l,n+1} (k_{\perp, j},m = 0) = \Psi_{{\rm Bulk},l,n+1} (k_{\perp, j},m = M)\\ &= \mathcal{A} \,\Psi_{l,n+1}(k_{\perp, j}) + \mathcal{B} \, \Psi_{l,n+1}(-k_{\perp, j}) = 0 \\
&\Rightarrow \mathcal{B}/\mathcal{A} = - \Psi_{l,n+1}(k_{\perp, j})/\Psi_{l,n+1}(-k_{\perp, j}),
\end{align*}
where $\Psi_{l,n+1}(k_{\perp, j})$ refers to the $(n+1)$th component of the $l$th Bloch wave $\Psi_{l}(k_\perp)$, which coincides with the amplitude of $\Psi_{l}(k_\perp)$ on the $B$ site. Setting $\mathcal{A} = 1$ and subsequently $\mathcal{B} = - \Psi_{l,n+1}(k_{\perp, j})/\Psi_{l,n+1}(-k_{\perp, j})$, we thus find the following solution for the generalized bulk state:
\begin{align}
&\Psi_{{\rm Bulk},l,i} (k_{\perp, j},m) = \nonumber \\
&e^{ik_{\perp, j} m}\Psi_{l,i}(k_{\perp, j})-e^{- i k_{\perp, j} m}\frac{\Psi_{l;n+1}(k_{\perp, j})}{\Psi_{l;n+1}(-k_{\perp, j})}\Psi_{l,i}(-k_{\perp, j}). \label{eq:bulksolutions}
\end{align}

It follows that Eq.~(\ref{eq:bulksolutions}) must also be solutions for the open chain with $M$ $A$ motifs and $(M-1)$ $B$ sites. Moreover, for $1\leq j,j'\leq M-1$, we find that these waves are orthogonal as long as $E_l(k_{\perp, j})\neq E_l(k_{\perp, j'})$. We have thus found $\left(M-1\right)(n+1)$ standing wave solutions. The open chain, however, has $M n+M-1=M(n+1)-1$ degrees of freedom, such that there are $n$ states unaccounted for. These missing states exactly correspond to the solutions in Eq.~(\ref{eqexactsolution}).

In the presence of the spectral symmetry, $E(k_\perp) = E(-k_\perp)$, the solutions in Eq.~(\ref{eqexactsolution}) are thus \emph{unique} in the sense that \emph{no other boundary modes} exist. 
If the spectral symmetry is broken, i.e., $E(k_\perp) \neq E(-k_\perp)$, the solutions in Eqs.~(\ref{eqexactsolution}) and (\ref{eqexactsolr}) remain valid and the boundary modes persist, while the solutions in Eq.~(\ref{eq:bulksolutions}) no longer exist. 

The construction presented here is applicable to a broad range of systems, and can be straightforwardly applied in multi-dimensional cases by requiring $E(\vec{k}_{||},k_\perp)=E(\vec{k}_{||},-k_\perp)$. In one-dimensional systems, this spectral symmetry is generated by mirror, inversion or time-reversal symmetry, whereas in higher-dimensional models only mirror symmetry leads to the desired spectral behavior. One might naively expect that other symmetries in two and three dimensions, e.g., inversion or time-reversal symmetry, would lead to a similar approach. This is prevented, however, by the fact that now the internal unit-cell labeling in the $A$ motifs starts to play a role and the unit cells in the $A$ motifs that are related to each other via these symmetries are not coupled to each other via the same site in $B$. The approach to find all bulk bands is, however, not destroyed by the presence of symmetries such as inversion or time reversal, \emph{as long as} the desired spectral mirror symmetry is also present. Note, moreover, that the only requirement we impose is that the spectrum has to exhibit the mirror symmetry, while it is not necessary that the Hamiltonian itself is also mirror symmetric. Indeed, in Sec.~\ref{sect:ci}, we present an example closely related to the Haldane model, and find that the spectrum has the desired symmetry while the Hamiltonian does not.

\subsection{Uniqueness} \label{sec:uniqueness}
In the previous sections, we found $n$ boundary-mode solutions [cf. Eq.~(\ref{eqexactsolution})], and showed they are unique in the case of a spectral mirror symmetry. However, even in the absence of such a spectral symmetry, it can be shown that no additional boundary states exist on the family of lattices that we study [cf. Fig.~\ref{figurechainmodels}(c)]. Indeed, the results by Dwivedi and Chua derived for transfer matrices \cite{dwivedichua} imply that in our models all boundary states are exponentially decaying waves with only a single exponent, i.e., $\Psi(m)=r^m \Psi$, where $\Psi(m)$ denotes the wave function in the $m$th unit cell. For $1<m<M$, we thus find
\begin{align*}
H_{B\rightarrow A}\Psi_{B}(m-1)+H_A\Psi_{A}(m)+H_{A\leftarrow B}&\Psi_{B}(m) \\
&= E\Psi_A(m),
\end{align*}
such that
\begin{equation}
H_{B\rightarrow A} r^{m-1}\Psi_{B}+H_A r^{m}\Psi_{A}+H_{A\leftarrow B} r^m\Psi_{B} = E r^m \Psi_{A}. \label{eq:conditionexpstate}
\end{equation} 
Similarly, for $m=1$, the following holds:
\begin{align*}
& H_A\Psi_{A}(1)+H_{A\leftarrow B}\Psi_{B}(1) = E\Psi_{A}(1)\\
& \Rightarrow H_A\Psi_{A}+H_{A\leftarrow B}\Psi_{B} = E\Psi_{A},
\end{align*}
where we note an equivalent expression is retrieved for $m = M$. Plugging this into Eq.~(\ref{eq:conditionexpstate}), we immediately find $\Psi_B=0$. We thus show that \emph{all} exponentially decaying wave functions that may exist on our lattices have a disappearing amplitude on \emph{all} $B$ sites. The solutions in Eq.~(\ref{eqexactsolution}) are thus the \emph{only} boundary mode solutions that exist.

\subsection{Spin degree of freedom} \label{sec:soc}

We now end our general section on exact solutions by showing that the constraint of having a single degree of freedom on each $B$ motif can be relaxed. As an explicit example, we consider spin degree of freedom, and realize that in the most simple case we may restrict ourselves to models where the spin-up and down sectors are decoupled. These systems can be straightforwardly solved with our formalism, and indeed we consider such an example in the form of a two-dimensional $\mathcal{Z}_2$ insulator in Sec.~\ref{sec:twodti}. While these $\mathcal{Z}_2$ insulators can be seen as being equal to the sum of two time-reversed Chern-insulator copies, their three-dimensional cousins cannot be interpreted as consisting of two separable systems, and spin-mixing terms generated by SOC are explicitly necessary to generate a three-dimensional topological insulator. These terms, however, may render the solutions to $r_i$ ambiguous.

There are, nevertheless, only two requirements that need to be satisfied for our formalism to be applicable also to this setting. Firstly, the eigenstates on the $A$ motifs need to have a well-defined spin, which can be established with a Hamiltonian of the form
\begin{align*}
H_A(k_{||})&=H_{A,SOC}(k_{||})\otimes \vec{d}(k_{||})\cdot\vec{\tau}+H_{A,0}\otimes \tau_0,
\end{align*}
where $\vec{\tau}$ is the vector of Pauli matrices acting in spin space, and the spin-mixing terms are thus captured by the first term. Secondly, the coupling between the $A$ and $B$ motifs has to be invariant under spin rotations, such that generally
\begin{align*}
H_{A\leftarrow B}&=H_{A\leftarrow B,0}\otimes \tau_0 + H_{A\leftarrow B,z}\otimes \tau_z, \\
H_{B\rightarrow A}&=H_{B\rightarrow A,0}\otimes \tau_0 + H_{B\rightarrow A,z}\otimes \tau_z.
\end{align*}
This means that for a given $k_{||}$, we have two decoupled spins polarized in the $\pm \vec{d}(k_{||})$ directions, and we can solve the boundary modes of this system using our method, where Eq.~(\ref{eqexactsolr}) for finding $r_i(k_{||})$ is modified to
\begin{equation}
r_i(k_{||}) = -\frac{\left(H_{A\leftarrow B}^\dagger\right)^{i,i}}{\left(H_{B\rightarrow A}^\dagger\right)^{i,i}}. \label{eqgenexprrextra}
\end{equation}
The dependence of the spin-polarization on $k_{||}$ is particularly important in the case of the three-dimensional topological insulator, because it enables us to obtain exact solutions for the surface Dirac cones as we explicitly show in an example in Sec.~\ref{sect:threedti}.

\section{Explicit examples} \label{sec:examples}

\begin{figure}[t]
  \centering
  \includegraphics[scale=0.245]{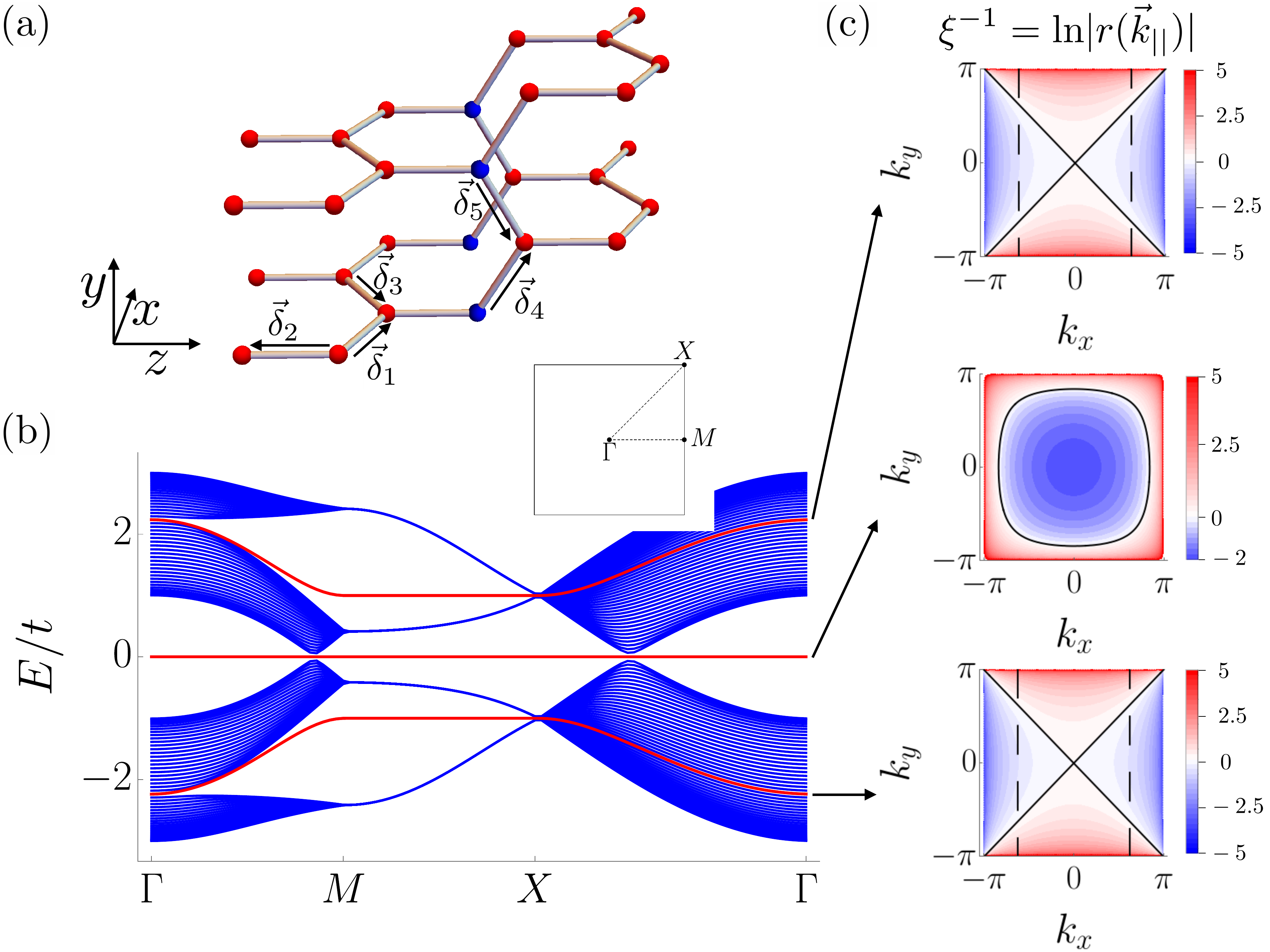}
  \caption{(a) Hyperhoneycomb lattice with the $A$ and $B$ motifs in red and blue, respectively, and the nearest-neighbor vectors $\vec{\delta}_i$ indicated explicitly. (b) Energy spectrum for the model in Eq.~(\ref{eq:hyperhoneyham}) for $M = 25$ with the bulk bands in blue and the states corresponding to the exact solution in red. (c) The inverse localization length $\xi^{-1} = \textrm{ln}|r(\vec{k})|$ for each of the three red bands with the black, solid line corresponding to $\xi^{-1} = 0$ and the black, dashed lines in the upper and lower panel are energy contours for $E = 1.5$ and $-1.5$, respectively, of the corresponding red bands.}
  \label{fighyperhoney}
\end{figure}

In this section, we present five explicit examples to showcase the formalism developed in the previous section. We start in Sec.~\ref{sect:nodallinesm} by considering a nearest-neighbor model on the three-dimensional hyperhoneycomb lattice, which realizes a nodal-line semimetal and show that we can find the exact solution for the drumhead state as well as two Fermi-arc-like states. We continue by considering a variation to the Haldane model \cite{haldane} in Sec.~\ref{sect:ci}, which we call the pseudo-Haldane model, and show that we can diagonalize the whole system by making use of the spectral symmetry $E(k_x, k_y) = E(k_x, -k_y)$ of the Bloch Hamiltonian to find the full spectrum. Next, we introduce spin-orbit coupling in Sec.~\ref{sec:twodti} by taking two time-reversed copies of the psuedo-Haldane model by which we realize a pseudo-Kane-Mele model \cite{kaneandmele,kaneandmeletwo}, which is a two-dimensional $\mathcal{Z}_2$ topological insulator. Again, we can solve the whole system exactly, and we find exact solutions to the helical edge states. In Sec.~\ref{sect:threedti}, we allow for spin mixing terms inside the $A$ motif and show that we can solve for the surface Dirac modes by studying a three-dimensional topological insulator on the diamond lattice. Lastly, we treat the two-dimensional Lieb lattice with open boundary conditions in two directions in Sec.~\ref{sec:lieb_lattice}, and show we can solve the full spectrum.

\subsection{Nodal-line semimetal} \label{sect:nodallinesm}

We start by considering the three-dimensional hyperhoneycomb lattice shown in Fig.~\ref{fighyperhoney}(a). This lattice has four sites in the unit cell, and we group them such that each $A$ [red in Fig.~\ref{fighyperhoney}(a)] and $B$ (blue) motif contains three and one sublattice site, respectively, and we do not consider additional degrees of freedom such that $n=3$. The nearest-neighbor vectors $\vec{\delta}_i$ are explicitly indicated in Fig.~\ref{fighyperhoney}(a), and read $\vec{\delta}_1 = (\sqrt{3}, \, 0,\, 1)/2$, $\vec{\delta}_2 = (0, \, 0,\, -1)$, $\vec{\delta}_3 = (-\sqrt{3}, \, 0,\, 1)/2$, $\vec{\delta}_4 = (0, \, \sqrt{3}, \, 1)/2$, $\vec{\delta}_5 = (0, \, -\sqrt{3}, \, 1)/2$, and the lattice vectors are defined as $\vec{a}_1 = \vec{\delta}_1-\vec{\delta}_3$, $\vec{a}_2 = \vec{\delta}_4-\vec{\delta}_5$, and $\vec{a}_3 = \vec{\delta}_3-2\vec{\delta}_2 + \vec{\delta}_4$. The Bloch Hamiltonian for this model is given in Eq.~(\ref{eqgeneralblochham}) with
\begin{align}
&H_A = \begin{pmatrix}
0 & t & 0 \\
t & 0 & t (1 + \textrm{e}^{i k_1}) \\
0 & t (1 + \textrm{e}^{-i k_1}) & 0
\end{pmatrix}, \qquad H_B = 0, \nonumber \\
&H_{A \leftarrow B} = \begin{pmatrix}
0 \\
0 \\
t
\end{pmatrix}, \qquad H_{B \rightarrow A} = \begin{pmatrix}
2 t \,  \textrm{e}^{-\frac{i}{2} k_1} \,\textrm{cos}(k_2/2) \\
0 \\
0
\end{pmatrix}, \nonumber \\
& k_\perp = 3 k_3 - k_2 + k_1, \label{eq:hyperhoneyham}
\end{align}
where $k_i \equiv \vec{k} \cdot \vec{a}_i$ and $t$ is the nearest-neighbor hopping parameter. Assuming open boundary conditions in $z$, and terminating with $A$ motifs on either end such that there is a bearded edge at $m = 1$ and a zigzag edge at $m=M$, we find the Hamiltonian $\hat{H} =\Psi^\dagger H \Psi$ with $\Psi^\dagger = (\psi_{A,1}^\dagger, \, \psi_{B,1}^\dagger, \, \psi_{A,2}^\dagger, \, \ldots , \, \psi_{B,M-1}^\dagger, \, \psi_{A,M}^\dagger)$, where $\alpha$ and $m$ in $\psi_{\alpha,m}^\dagger$ label the motif and unit cell, respectively, and
\begin{equation}
H = \begin{pmatrix}
H_A & H_{A \leftarrow B} & 0 & 0 & 0 \\
H_{A \leftarrow B}^\dagger & H_B & H_{B \rightarrow A}^\dagger & 0 & 0 \\
0 & H_{B \rightarrow A} & H_A & \cdots & 0 \\
0 & 0 & \vdots & \ddots & H_{B \rightarrow A}^\dagger \\
0 & 0 & 0 & H_{B \rightarrow A} & H_A
\end{pmatrix}. \label{eqopenham}
\end{equation}
\begin{figure}[t]
  \centering
  \includegraphics[scale=0.38]{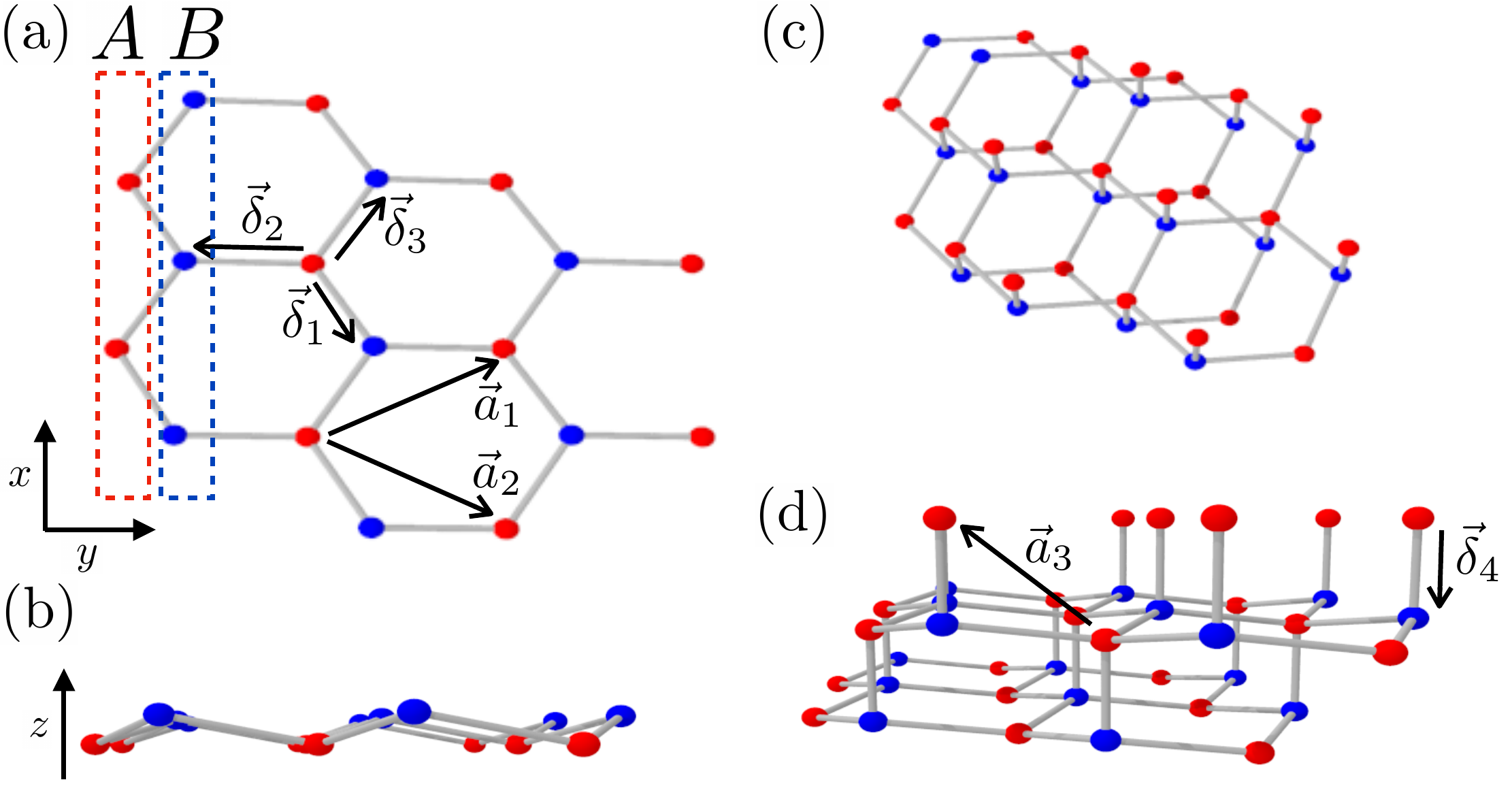}
  \caption{(a) The honeycomb lattice with the $A$ and $B$ motifs corresponding to periodic chains shown by the red and blue dashed boxes, respectively, with a zigzag and bearded edge. The nearest-neighbor and lattice vectors $\vec{\delta}_i$ and $\vec{a}_i$ are indicated with arrows. (b) Side view of the buckled honeycomb lattice. [(c) and (d)] Top and side view of the diamond lattice with $(111)$ surfaces, where the honeycomb lattices are buckled [cf. (b)], with the lattice vectors as defined in (a) and the additional vectors $\vec{\delta}_4$ and $\vec{a}_3$ in (d).}
  \label{fighoneycomblatt}
\end{figure}
The band spectrum for this model is shown in Fig.~\ref{fighyperhoney}(b), where we have rescaled $k_x \rightarrow k_x/\sqrt{3}$. As $H_A$ is not written in a diagonal form, the exact solutions for the wave functions are given in Eq.~(\ref{eqexactsolrnondiag}) with $r_i (\vec{k}_{||})$ derived from Eq.~(\ref{eqexactsolrextra})
\begin{equation*}
r_i (\vec{k}_{||})=-\frac{\phi_{i,3}}{2\,  \textrm{e}^{\frac{i}{2} k_1} \,\textrm{cos}(k_2/2) \phi_{i,1}},
\end{equation*}
where $\phi_{i,\bar{n}}$ are the eigenfunctions of $H_A$. The eigenvalues $E_i$ associated with the wave functions in Eq.~(\ref{eqexactsolrnondiag}) are equivalent to the eigenvalues of $H_A$, i.e., $H_A \phi_i = E_i \phi_i$, with $E_i = 0, \, \pm t \textrm{e}^{-i k_1} \sqrt{\textrm{e}^{i k_1} + 3 \textrm{e}^{2 i k_1} + \textrm{e}^{3 i k_1}}$, and are shown in red in Fig.~\ref{fighyperhoney}(b). Plotting the inverse localization length $\xi^{-1}_i(\vec{k}_{||}) = \textrm{ln}|r_i (\vec{k}_{||})|$, such that $\xi^{-1}_i(\vec{k}_{||}) < 0$ ($>0$) when $|r_i(\vec{k}_{||})|<1$ ($>1$), for the flat band as shown in the middle panel of Fig.~\ref{fighyperhoney}(c), we clearly see that it corresponds to a drumhead state, and that the bulk-gap closing at which the band switches localization forms a closed loop (in black) revealing this is a nodal-line semimetal. The other two red bands also switch localization as shown in the upper and lower panel of Fig.~\ref{fighyperhoney}(c). We have plotted energy contours (black, dashed lines) on top of the inverse localization length such that it is clearly visible that, following these contours, both these states switch surfaces. These bands thus have a Fermi-arc-like character in the sense that they are broken and live on opposite surfaces.

\begin{figure*}[t]
  \centering
  \includegraphics[scale=0.5]{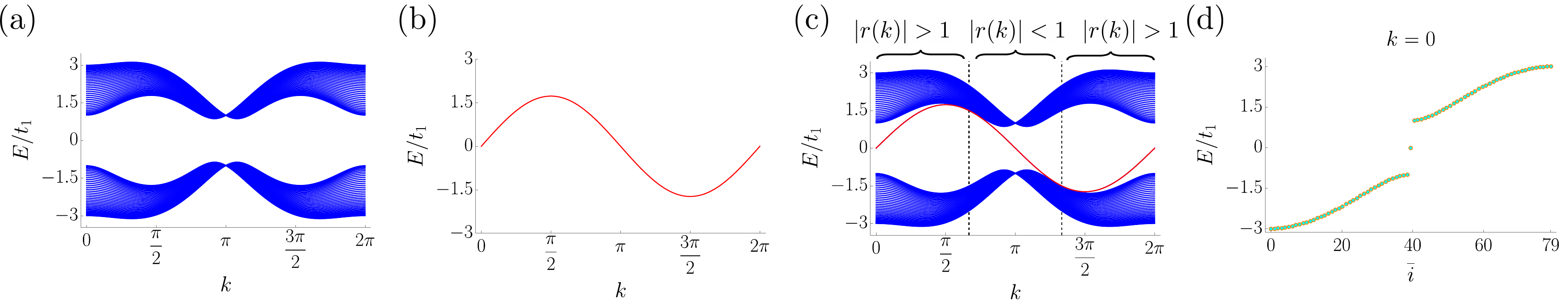}
  \caption{Energy spectrum of the pseudo-Haldane model in Eq.~(\ref{eqblochhampseudohaldane}) for $M = 40$ and $t_2/t_1 = \sqrt{3}$ with (a) the bulk bands computed using $E = E_l(2 \pi j /M)$ with $j = 1, \ldots, (M-1)$, (b) the chiral mode is given by Eq.~(\ref{eqexactsolution}), and (c) the full band spectrum from joining the bulk and boundary modes. The black dashed line correspond to solutions to $|r(k)| = 1$, and the regions with $|r(k)| < 1$ and $|r(k)| > 1$ are explicitly indicated. (d) The energies obtained via ED and via our analytical method are shown in orange and cyan, respectively, for the cut $k=0$ with $\bar{i}$ the degrees of freedom of the full system, and overlap up to machine precision.}
  \label{figenergyhoneyiqh}
\end{figure*}

\subsection{Chern insulator} \label{sect:ci}

We continue by introducing a variation to the Haldane model \cite{haldane}, which we call the pseudo-Haldane model, which has sublattice degree of freedom and realizes a Chern-insulating phase with chiral, one-dimensional edge modes on the two-dimensional honeycomb lattice shown in Fig.~\ref{fighoneycomblatt}(a). In this case, the $A$ and $B$ motifs simply correspond to periodic chains consisting of the $A$ and $B$ sublattices of the honeycomb lattice only. Here we find the exact solution corresponding to the chiral edge modes, as was previously demonstrated for a lattice with chiral hinge states in Ref.~\onlinecite{kunstvmiertbergholtz}, and also all solutions for the bulk states, which is a new result. The nearest-neighbor vectors $\vec{\delta}_i$ and lattice vectors $\vec{a}_i$ are explicitly indicated in Fig.~\ref{fighoneycomblatt}(a), and read $\vec{\delta}_1 = (-\sqrt{3}, \, 1,\, 0)/2$, $\vec{\delta}_2 = (0, \, -1,\, 0)$, $\vec{\delta}_3 = (\sqrt{3}, \, 1,\, 0)/2$, $\vec{a}_1 = \vec{\delta}_3-\vec{\delta}_2$ and $\vec{a}_2 = \vec{\delta}_1-\vec{\delta}_2$. The Bloch Hamiltonian is given by Eq.~(\ref{eqgeneralblochham}) with
\begin{align}
&H_A = - H_B = t_2 \, {\rm sin}(k_2 - k_1), \quad H_{B \rightarrow A} = t_1 \textrm{e}^{\frac{i}{2} (k_2 - k_1)}, \nonumber \\
& H_{A \leftarrow B} = t_1 (1 + \textrm{e}^{i (k_2 - k_1)}), \qquad k_\perp = - (k_1 + k_2)/2, \label{eqblochhampseudohaldane}
\end{align}
where $k_i \equiv \vec{k} \cdot \vec{a}_i$, and $t_1$ ($t_2$) are the (next-)nearest-neighbor hopping parameters. We point out that this pseudo-Haldane model differs from the original Haldane model in that we have turned off all the next-nearest-neighbor hopping terms with a $k_y$-dependence, such that there is no long-range hopping connecting different $A$ motifs. The eigenvalues of the Bloch Hamiltonian have the spectral symmetry $E(k_x, k_y) = E(k_x, - k_y)$ such that we can use the method in section~\ref{sect:fulldiag} to fully diagonalize this model. Note that this model has additional spectral symmetries, $E(k_x, k_y) = E(-k_x, k_y) = E(-k_x, - k_y)$, which are in principle not required for the solvability of the model.

Assuming open boundary conditions in $y$ while keeping periodicity in $x$, and terminating with $A$ motifs on both sides, such that there is a zigzag and bearded edge at $m = 1$ and $m=M$, respectively, we find the Hamiltonian given in Eq.~(\ref{eqopenham}) with the entries given in Eq.~(\ref{eqblochhampseudohaldane}). The bulk bands computed using $E = E_l(k_x, 2 \pi j /M)$ with $j = 1, \ldots, M-1$ are shown in Fig.~\ref{figenergyhoneyiqh}(a), while the chiral band with energy $H_A = t_2 \, {\rm sin}(k)$ with $k = k_1 - k_2 = \sqrt{3} k_x$ is shown in Fig.~\ref{figenergyhoneyiqh}(b). Combining these two spectra, we find Fig.~\ref{figenergyhoneyiqh}(c), which corresponds exactly to results obtained via exact diagonalization (ED). Indeed, in Fig.~\ref{figenergyhoneyiqh}(d) we compare the energies obtained via ED (in orange) and to those obtained analytically (in cyan) for a specific cut in $k$, and find that they overlap up to machine precision. The wave function corresponding to this chiral mode is given by Eq.~(\ref{eqexactsolution}) with $r(k)$ defined in Eq.~(\ref{eqexactsolr}) such that we find $r(k) = -2 \, \textrm{cos}(k/2)$, and the mode necessarily attaches to the bulk when $k = \pm 2 \pi/3 \, \textrm{mod} \, 2 \pi$, i.e., when $|r(k)| = 1$ [black dashed lines in Fig.~\ref{figenergyhoneyiqh}(c)]. Moreover, from the specific form of $r(k)$ we deduce that the right (left) mover is localized to the $A$ lattice at the edge $m = M$ ($m = 1$).

\begin{figure}[b]
  \centering
  \includegraphics[scale=0.4]{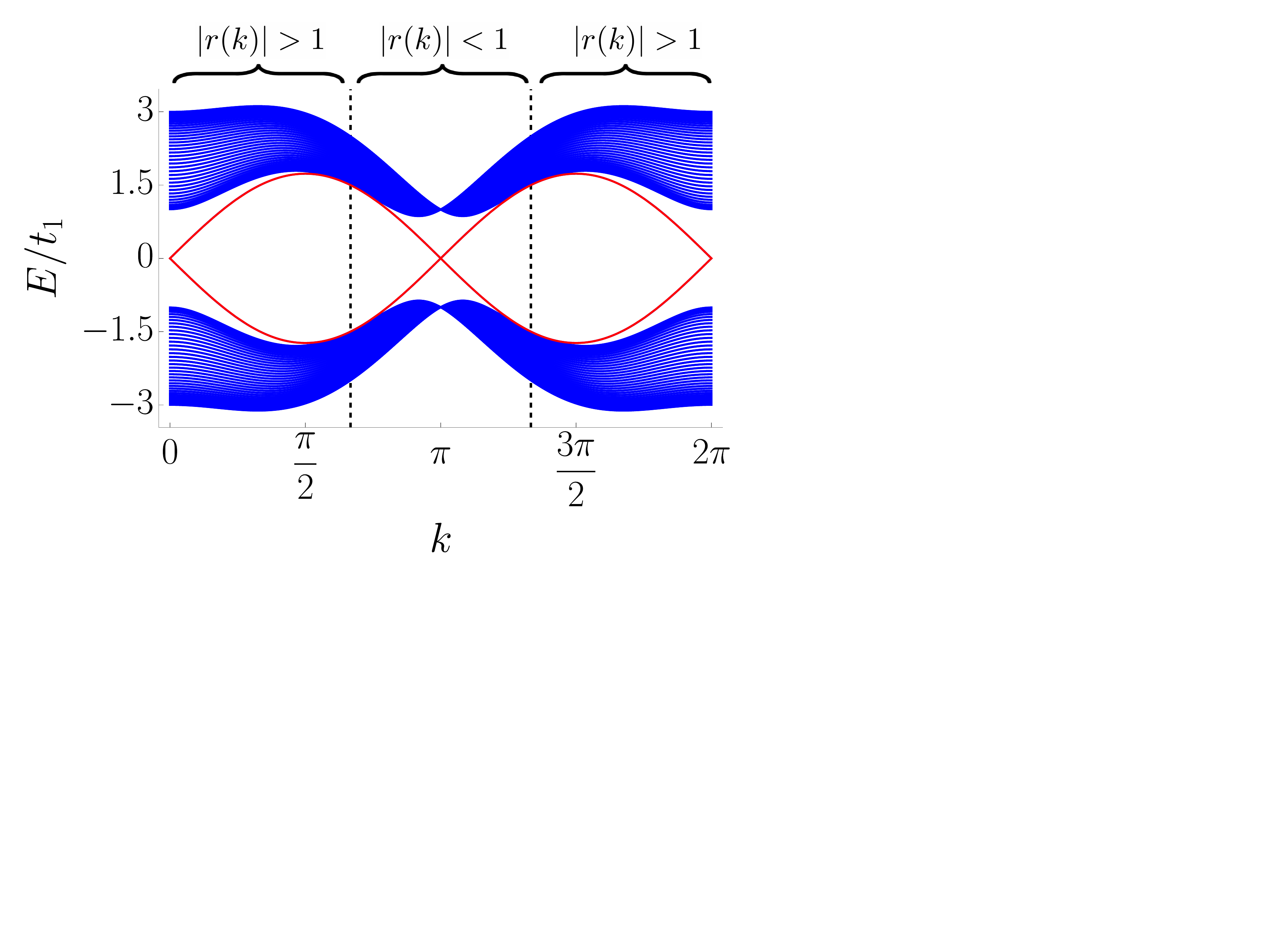}
  \caption{Energy spectrum of the pseudo-Kane-Mele model in Eq.~(\ref{eqblochhampseudokanemele}) with the chiral mode given by Eq.~(\ref{eqexactsolution}) in red with $M = 40$, and $t_2/t_1 = \sqrt{3}$. The black dashed line correspond to solutions to $|r(k)| = 1$, and the regions with $|r(k)| < 1$ and $|r(k)| > 1$ are explicitly indicated.}
  \label{figenergyhoneysqh}
\end{figure}

\begin{figure*}[t]
  \centering
  \includegraphics[scale=0.5]{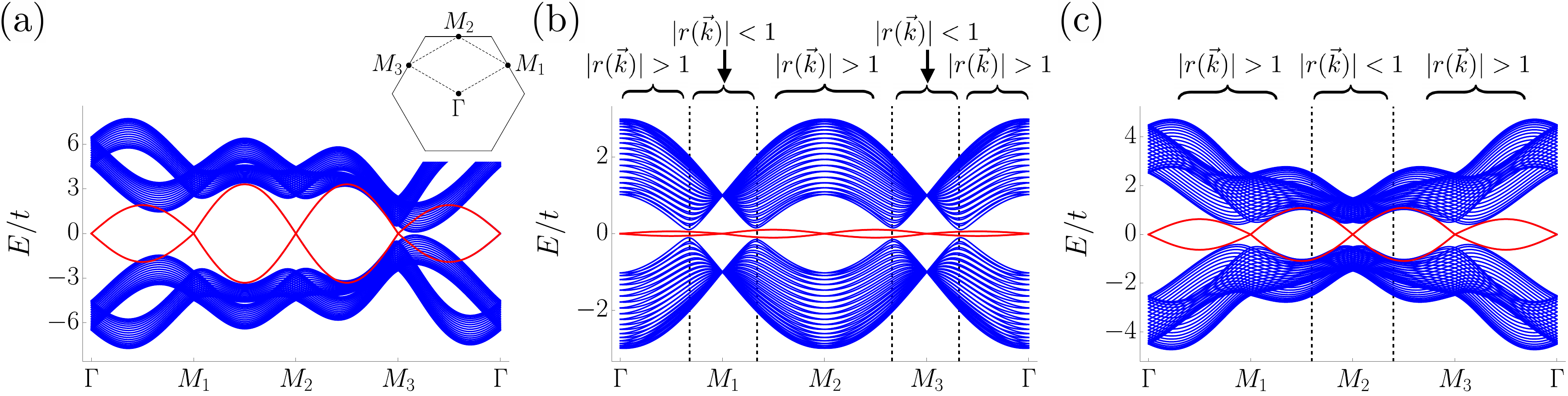}
  \caption{Energy spectrum of the three-dimensional topological insulator model described in Eq.~(\ref{eqdiamondham}) with the bulk bands in blue and the exact solution in red for $M = 20$, (a) $\lambda/t = 0.676$ and $\delta t_1/t = 5/2$, (b) $\lambda/t = 0.064/3$ and $\delta t_2/t = -1$ and (c) $\lambda/t = 2/9$ and $\delta t_2/t = 0.5$. The Brillouin zone is shown as an inset in (a). The black dashed lines in (b) and (c) correspond to solutions to $|r(\vec{k})| = 1$, and the regions where $|r(\vec{k})| < 1$ and $|r(\vec{k})| > 1$ are explicitly indicated. In (a), $|r(\vec{k})| > 1$ for all $\vec{k}$.}
  \label{figdiamondlatt}
\end{figure*}

\subsection{Two-dimensional topological insulator} \label{sec:twodti}

We now increase the number of degrees of freedom on the $A$ and $B$ motifs by introducing spin, and study a modified version of the Kane-Mele model \cite{kaneandmele,kaneandmeletwo}, which we call the pseudo-Kane-Mele model. This system explicitly includes SOC terms, and realizes the quantum spin Hall effect on the graphene lattice, which features helical states on its edges. The Bloch Hamiltonian is given by Eq.~(\ref{eqgeneralblochham}) with
\begin{align}
&H_A = - H_B = t_2 \, \textrm{sin}(k_2 - k_1) \tau_z, \,\, H_{B \rightarrow A} = t_1 \textrm{e}^{\frac{i}{2}(k_2 - k_1)} \tau_0, \nonumber \\
& H_{A \leftarrow B} = t_1 (1 + e^{i (k_2 - k_1)}) \tau_0, \qquad k_\perp = -(k_1 + k_2)/2, \label{eqblochhampseudokanemele}
\end{align}
with $\tau_i$ the Pauli matrices acting in spin space. Note that the terms proportional to $t_2$ now play the role of mirror-symmetric SOC \cite{kaneandmele,kaneandmeletwo}. This model is closely related to the pseudo-Haldane model in Eq.~(\ref{eqblochhampseudohaldane}) in the sense that the pseudo-Kane-Mele model realizes two time-reversed copies of the pseudo-Haldane model. There is only SOC inside each $A$ and $B$ motif, which leaves the spin up and down sectors decoupled. This means that, as in the previous section, we again have a spectral symmetry $E(k_x, k_y) = E(k_x, -k_y)$, and we can diagonalize the full model.

Taking open boundary conditions in $y$, the Hamiltonian is given in Eq.~(\ref{eqopenham}) with the entries in Eq.~(\ref{eqblochhampseudokanemele}), and the accompanying band spectrum is shown in Fig.~\ref{figenergyhoneysqh}. The exact wave-function solutions are given in Eq.~(\ref{eqexactsolution}) with the eigenenergies $E_1 (k) = - E_2 (k) = t_2 \, \textrm{sin}(k)$ (red bands in Fig.~\ref{figenergyhoneysqh}) the eigenvalues of $H_A$, where $k = k_1 - k_2 = \sqrt{3} k_x$. We use Eq.~(\ref{eqgenexprrextra}) to find $r_i(k)$ such that $r(k) = r_1 (k) = r_2 (k) = -2 \, \textrm{cos}(k/2)$ as before. We have indicated the value of $|r(k)|$ explicitly in Fig.~\ref{figenergyhoneysqh}, and we see that two helical states localize to both edges.

\subsection{Three-dimensional topological insulator} \label{sect:threedti}
Here, we study a three-dimensional extension of the previous model in the form of a strong topological insulator in the diamond lattice \cite{kanemelestrongti}. We investigate the lattice with $(111)$ surfaces, such that the model consists of stacked, buckled honeycomb lattices as shown in Fig.~\ref{fighoneycomblatt}(b)-\ref{fighoneycomblatt}(d). Each of the triangular $A$ [red in Fig.~\ref{fighoneycomblatt}(b)] and $B$ (in blue) sublattices of a buckled honeycomb sheet makes up the $A$ and $B$ motifs of the full model [cf. Figs.~\ref{fighoneycomblatt}(c)-\ref{fighoneycomblatt}(d)]. The nearest-neighbor vectors now read $\vec{\delta}_1 = (-\sqrt{2}/\sqrt{3}, \, \sqrt{2}/3,\, 1/3)$, $\vec{\delta}_2 = (0, \, -2\sqrt{2}/3,\, 1/3)$, $\vec{\delta}_3 = (\sqrt{2}/\sqrt{3}, \, \sqrt{2}/3,\, 1/3)$ and $\vec{\delta}_4 = (0, \, 0, \, -1)$, while the next-nearest-neighbor vectors $\vec{a}_i$ are defined as in Sec.~\ref{sect:ci} with the additional vector $\vec{a}_3 = \vec{\delta}_2 - \vec{\delta}_4$. We adopt the Hamiltonian introduced by Fu \emph{et al.} in Ref.~\onlinecite{kanemelestrongti}, such that the Bloch Hamiltonian is given by Eq.~(\ref{eqgeneralblochham}) with
\begin{align}
& H_A = - H_B = \lambda \left\{ -\sqrt{2} \right[\textrm{sin} \left(k_1 \right) + \textrm{sin} \left(k_2 \right) \left] \tau_x \right. \nonumber \\
& \left. + \frac{\sqrt{2}}{\sqrt{3}} \left[\textrm{sin} \left(k_1 \right) - \textrm{sin} \left(k_2 \right) + 2 \textrm{sin} \left(k_1-k_2 \right) \right] \tau_y \right. \nonumber \\
& \left. + \frac{4}{\sqrt{3}} \left[\textrm{sin} \left(k_1 \right)  -\textrm{sin} \left(k_2 \right) - \textrm{sin} \left(k_1-k_2 \right)\right] \tau_z  \right\} , \nonumber \\
&H_{A \leftarrow B} = [t_1 + t_2 \textrm{e}^{i k_2} + t_3\textrm{e}^{-i (k_1-k_2)}]\tau_0, \nonumber \\
&H_{B \rightarrow A} = t_4 \,  \textrm{e}^{i (2 k_2 - k_1)/3} \tau_0, \quad k_\perp = - (k_1 + k_2)/3 - k_3, \label{eqdiamondham}
\end{align} 
with $k_i \equiv \vec{k} \cdot \vec{a}_i$, $t_i = t + \delta t_i$ the nearest-neighbor hopping parameter, $\lambda$ the next-nearest-neighbor SOC term, and $\tau_i$ acting in spin space as before. As opposed to the model in Ref.~\onlinecite{kanemelestrongti}, we only include SOC inside each $A$ ($B$) layer and not between $A$ ($B$) layers. Taking open boundary conditions in $z$, the Hamiltonian is given in Eq.~(\ref{eqopenham}) with the entries in Eq.~(\ref{eqdiamondham}) and the energy spectrum is shown in Fig.~\ref{figdiamondlatt}. The exact wave-function solutions are given in Eq.~(\ref{eqexactsolrnondiag}), and the eigenvectors $\phi_i(\vec{k})$ and eigenvalues $E_i(\vec{k})$ (red bands in Fig.~\ref{figdiamondlatt}) are found by diagonalizing $H_A$. We thus find that there are always four (surface) Dirac cones in the system when $\lambda \neq 0$. As $H_{A \leftarrow B}$ and $H_{B\rightarrow A}$ are diagonal matrices, we find that, even though $H_A$ is not written in its diagonal form, we can use Eq.~(\ref{eqgenexprrextra}) to find $r_i(\vec{k})$
\begin{equation*}
r_1 (\vec{k}) = r_2 (\vec{k}) = - \frac{t_1 + t_2 \textrm{e}^{-i k_2} + t_3\textrm{e}^{i (k_1-k_2)}}{t_4 \,  \textrm{e}^{i (2 k_2 - k_1)/3}}.
\end{equation*}
\begin{figure*}[t]
  \centering
  \includegraphics[scale=0.5]{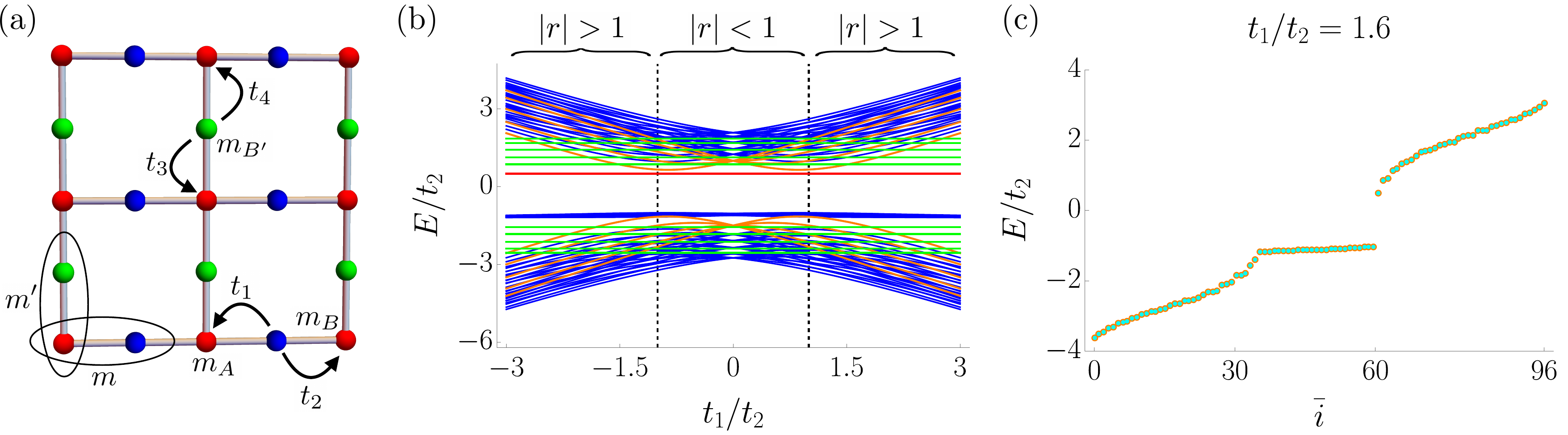}
  \caption{(a) Schematic Lieb lattice with the $A$, $B$, and $B'$ motifs in red, blue, and green, respectively, and the Hamiltonian explicitly depicted. The unit cells of each $AB$ chain and $AB'$ chains are labeled by $m$ and $m'$, respectively. (b) Band spectrum as a function of $t_1/t_2$ with $M = M' = 6$, $m_A/t_2 = 0.5$, $m_B/t_2 = -1$, $m_{B'} /t_2 = -1.2$, $t_3/t_2 = 0.7$, and $t_4/t_2 = 1.4$ with the bulk bands in blue, the $AB$ and $AB'$ edge bands in orange and green, respectively, and the corner energy in red. $r' = -1/2$ for this model, the black dashed lines correspond to $|r| = 1$, and the regions where $|r|<1$ and $|r|>1$ are explicitly indicated. (c) The energies obtained via ED and via our analytical method are shown in orange and cyan, respectively, for the cut $t_1/t_2=1.6$ with $\bar{i}$ the degrees of freedom of the full system, and overlap up to machine precision.}
  \label{figlieblatt}
\end{figure*}
Depending on the choice of $t_i$, $r(\vec{k})$ switch between $|r(\vec{k})|<1$ and $|r(\vec{k})|>1$ as a function of $\vec{k}$. This is indeed shown in Fig.~\ref{figdiamondlatt}, where we find that the localization of the Dirac cones changes depending on the choice for $t_i$. Indeed, when we turn on $\delta t_1$ only, we find that the system is a trivial insulator as the four cones are localized to the same surface [cf. Fig.~\ref{figdiamondlatt}(a)]. However, when we turn on $\delta t_2$ instead, we find we can realize both a weak [Fig.~\ref{figdiamondlatt}(b)] and strong topological insulator [Fig.~\ref{figdiamondlatt}(c)]. In the first case, this phase is found by tuning $\delta t_2$ such that the hopping $t_2 = 0$, which effectively reduces the diamond lattice into a stacking of decoupled honeycomb lattices with edges in the open surfaces. Each of these lattices realizes a quantum-spin Hall insulator with helical edge states, such that we find two Dirac cones on either of the surfaces. In Fig.~\ref{figdiamondlatt}(c), we tune $\delta t_2$ such that $t_2 > t_i , \, i = 1, 3,4$, and we find that one Dirac cone localizes to the bottom surface $m=1$ whereas the three other cones localize to the top surface $m = M$. We thus recover a strong topological insulator.

We note that even though in the case of the trivial insulator [cf. Fig.~\ref{figdiamondlatt}(a)] the surface state enters the bulk bands, it does not hybridize with the bands at the same energy. Indeed, the degeneracy in energy in this case is accidental and we could add local operators to the Hamiltonian to separate the solvable surface state from the continuous bulk spectrum. This can always be done whenever $|r(k)| \neq 1$, i.e., whenever Eq.~(\ref{eqexactsolution}) describes a surface state.

Lastly, we point out that if $\lambda = 0$, this model [cf. Eq.~(\ref{eqdiamondham})] reduces to a nodal-line semimetal with a doubly degenerate drumhead state. This is indeed already visible in Fig.~\ref{figdiamondlatt}(b), where the SOC is only turned on weakly. We have thus established a direct link between Dirac surface states and drumhead states. Moreover, we note that they are captured by the \emph{exact} same wave function, and only their eigenenergies are affected by the change in $\lambda$.

\subsection{Generalization to the Lieb lattice} \label{sec:lieb_lattice}

Lastly, we turn to the two-dimensional Lieb lattice shown schematically in Fig.~\ref{figlieblatt}(a). This lattice is a two-dimensional generalization of the chain in Fig.~\ref{figurechainmodels}(c), and consists of $A$ (in red), $B$ (in blue), and $B'$ (in green) motifs with each one degree of freedom. The real-space Hamiltonian is depicted in Fig.~\ref{figlieblatt}(a), and reads
\begin{align*}
\hat{H}&=\sum_{m,m'} \sum_{\alpha\in\{A,B,B'\}} m_\alpha c^\dagger_{\alpha,m} c_{\alpha,m} \nonumber\\
&+\sum_{m,m'}(t_1 \, c^\dagger_{A,m,m'} c_{B,m,m'} +t_2 \, c^\dagger_{A,m+1,m'} c_{B,m,m'} \nonumber\\
&+t_3 \, c^\dagger_{A,m,m'} c_{B',m,m'} +t_4 \, c^\dagger_{A,m,m'+1} c_{B',m,m'} + H.c.).
\end{align*}
In Refs.~\onlinecite{kunstvmiertbergholtz, kunstvmiertbergholtz2}, we showed that
\begin{equation}
\ket{\psi} = \mathcal{N} \mathcal{N}' \sum_{m=1}^M \sum_{m'=1}^{M'} r^m r'^{m'} c^\dagger_{A,m}\ket{0}, \label{eq:cornerstate}
\end{equation}
with $r = -t_1/t_2$ and $r' = -t_3/t_4$, is an exact solution on this lattice with energy $E = m_A$, which corresponds to a corner state when $|r|\lessgtr1$ and $|r'|\lessgtr1$, while it represents an edge state localized to the bottom (top) edge formed by $A$ and $B$ motifs when $|r| = 1$ and $|r'|<1$ ($>1$), and localized to the left (right) edge formed by $A$ and $B'$ motifs when $|r'| = 1$ and $|r|<1$ ($>1$).

The Bloch Hamiltonian for this model reads $\hat{H}(\vec{k})=(c^\dagger_{A, \vec{k}},c^\dagger_{B, \vec{k}},c^\dagger_{B', \vec{k}})H(\vec{k})(c_{A, \vec{k}},c_{B, \vec{k}},c_{B', \vec{k}})^T $ with
\begin{align*}
H(\vec{k}) &=\begin{pmatrix}
m_A&t_1+t_2e^{-ik_x} & t_3+t_4e^{-ik_y}\\
t_1+t_2e^{ik_x} &m_B & 0 \\
t_3+t_4e^{ik_y}& 0 & m_{B'}
\end{pmatrix}.
\end{align*}
We find the eigenstates of this Bloch Hamiltonian have the spectral mirror symmetry $E(k_x,k_y) = E(-k_x,k_y) = E(k_x,-k_y)=E(-k_x,-k_y)$ regardless of our choice for parameters. We can also write the Hamiltonians corresponding to the edges formed by $A$ and $B$, and $A$ and $B'$ motifs as
\begin{align*}
H_{\textrm{edge},AB} (k_x) &= \begin{pmatrix}
m_A & t_1+t_2e^{-ik_x} \\
t_1+t_2e^{ik_x} & m_B
\end{pmatrix}, \\
H_{\textrm{edge},AB'} (k_y) &= \begin{pmatrix}
m_A & t_3+t_4e^{-ik_y} \\
t_3+t_4e^{ik_y} & m_{B'}
\end{pmatrix},
\end{align*}
respectively. In Fig.~\ref{figlieblatt}(b), we show the spectrum for this model with open boundary conditions in $x$ and $y$. In addition to the corner state solution (in red), whose eigenstate is given in Eq.~(\ref{eq:cornerstate}) and whose energy is corresponds to $m_A$, we find we can also accurately compute the remainder of the band spectrum using $E(\pi j/M, \pi j'/M')$ with $E(k_x,k_y)$ the eigenstates of the Bloch Hamiltonian, to find the bulk bands (in blue), and $E_{\textrm{edge},AB} (\pi j/M)$ and $E_{\textrm{edge},AB'} (\pi j'/M')$ to retrieve the edge bands (in orange and green, respectively), where $j = 1 , \ldots, M-1$ and $j' = 1 , \ldots, M'-1$. Indeed, in Fig.~\ref{figlieblatt}(c), we explicitly show the overlap of the results computed using ED (in orange) and those retrieved with the above description (in cyan) for a specific choice of parameters, and find an overlap up to machine precision.

\section{Discussion} \label{sec:discussion}

In this work, we address several open questions concerning the method we developed in Refs.~\onlinecite{kunsttrescherbergholtz, kunstvmiertbergholtz, kunstvmiertbergholtz2} to find exact boundary state solutions. We show that in the presence of a spectral mirror symmetry, which relates the open boundaries, we can solve for the complete eigensystem, and that our boundary states are unique, also in the absence of this symmetry. Moreover, we consider the possibility of including SOC in our models, and show that only weak conditions need to be satisfied for us to be able to still solve the model. We emphasize that to our knowledge no exact solutions have been found that capture drumhead surface state, nor has the close relation between Dirac surface states and doubly-degenerate drumhead states---they have \emph{the exact same wave function}---been established before. Moreover, to the best of our knowledge, we are the first to provide exact analytical solutions for the entire band spectrum of the two-dimensional Lieb lattice with open boundary conditions in two directions.

We note that other methods exist for finding exact wave-function solutions, such as the aforementioned transfer-matrix formalism \cite{leejoannopoulos, hatsugai, dwivedichua}, a rather involved method based on the generalization of Bloch's theorem \cite{alasecobaneraortizviola, cobaneraalaseortizviola}, a construction specifically catered to time-reversal-symmetry preserving systems \cite{duncanoehbergvaliente}, and a formalism making use of boundary Green's functions leading to effective boundary theories \cite{lopezsanchotworubio, umerski, pengbaovonoppen}. In case of the transfer-matrix approach, we point out that while transfer matrices specifically allow for the computation of bulk and boundary states, eigenvalues are treated on an equal footing with the other parameters in the model, and only solutions that are allowed to exist on the lattice come out as valid. This is in sharp contrast to the approach we have presented here, where the energy eigenvalues for the boundary states are computed directly from the Hamiltonian on the $A$ motif, $H_A$, while the bulk energies come about by solving the eigenvalue equation for the Bloch Hamiltonian $H(\vec{k}_{||},k_\perp)$. We, moreover, note that the approach for finding bulk modes presented in Ref.~\onlinecite{duncanoehbergvaliente} for systems with open boundary conditions in one direction carries some resemblance to our method in Sec.~\ref{sect:fulldiag}. Whereas the focus in Ref.~\onlinecite{duncanoehbergvaliente} is on time-reversal-symmetric models, we observe that the minimal generic requirement for using this method is in fact the presence of a spectral mirror symmetry relating the open boundaries.

Our work can presumably also be extended to other physical contexts where exact solutions are likely to be illuminating. One such interesting new area of application for our results in the field of non-Hermitian physics, relevant for a range of dissipative systems, in the context of topology, which has attracted much recent attention due to several exotic phenomena ranging from the breakdown of conventional bulk-boundary correspondence \cite{kunstedvardssonbudichbergholtz, lee} to the appearance of exceptional points, which correspond to a degeneracy both in the space of energies and the space of eigenvectors \cite{martinezalvarezbarriosvargasberdakinfaotorres}. The failure of conventional bulk-boundary correspondence is strongly related to a phenomenon dubbed the ``skin effect," where bulk states pile up at the boundary \cite{kunstedvardssonbudichbergholtz, martinezalvarezbarriosvargasberdakinfaotorres, kunstdwivedi, yaowang, yaosongwang, leethomale}. In this case, the continuum states thus behave as boundary modes, and a study of such bands following our method in Sec.~\ref{sect:fulldiag} is thus particularly interesting. Indeed, initial insight was recently enabled by insights from exact solutions in the simplest one and two-dimensional settings \cite{kunstedvardssonbudichbergholtz}. Our new way of obtaining the full spectrum analytically appears as particularly promising for understanding this rich class of systems better.

\acknowledgments
We would like to thank M. Trescher for related collaborations, and V. Dwivedi for useful discussions. F.K.K. and E.J.B. are supported by the Swedish research council (VR) and the Wallenberg Academy Fellows program of the Knut and Alice Wallenberg Foundation. G.v.M. is supported by the research program of the Foundation for Fundamental Research on Matter (FOM), which is part of the Netherlands Organization for Scientific Research (NWO).

\end{document}